\documentclass[aps,prd,twocolumn,showpacs,amsmath]{revtex4}
\usepackage[dvips]{color,graphicx}
\usepackage{amsfonts}
\usepackage{amssymb}
\usepackage{latexsym}

\newcommand{\dalm}{\kern1pt\vbox{\hrule height 0.9pt\hbox{\vrule width
0.9pt\hskip 2.5pt\vbox{\vskip 5.5pt}\hskip 3pt\vrule width 0.3pt}\hrule height
0.3pt}\kern1pt}

\newcommand{\lw}[1]{\smash{\lower2.ex\hbox{#1}}}

\begin{document}

%\thispagestyle{empty}

%<<<<<<<<<<<<< TITLE >>>>>>>>>>>>>>>%
\title{Einstein is Newton with space curved}
%<<<<<<<<<<<<< AUTHOR >>>>>>>>>>>>>>>%
\author{Naresh Dadhich\thanks{Electronic address:nkd@iucaa.ernet.in}}
\email{nkd@iucaa.ernet.in}
%<<<<<<<<<<<<< ADDRESS >>>>>>>>>>>>>>>%
\affiliation{Centre for Theoretical Physics, Jamia Millia Islamia, New Delhi~110025 \\
IUCAA, Post Bag 4, Pune~411007\\}
\date{\today}

%======================================%
%<<<<<<<<<<<<< ABSTRACT >>>>>>>>>>>>>>>%
%======================================%
\begin{abstract}
The two defining features of the Einsteinian gravity are that it is self interactive as well as it links universally to all particles including zero mass particles. In
the process of obtaining the Schwarzshild solution for gravitational field of a mass point, we would demonstrate how are these features 
incorporated? We would also show, unlike the Newtonian gravity, why the gravitational potential in the Schwarzshild solution can have its zero
only at infinity and nowhere else;i.e. it is determined absolutely. We also consider particle orbits to expose certain insightful and subtle points of concept and principle. 
\end{abstract}

%<<<<<<<<<<<<< PACS NUMBER >>>>>>>>>>>>>>>%
\pacs{04.20.-q, 04.20.Cv}

\maketitle

\section{Introduction}
The Newtonian gravity links to all massive particles and they attract each-other by the inverse square law force. However massless particle or
light remains unaffected. The Einsteinian gravity results from universalization of the Newtonian gravity. That is to include massless particles
in the gravitational interaction. This requirement uniquely asks for gravity to be described by spacetime curvature \cite{dad1}. Of course
universalization also means that energy distribution in any form including the energy of gravitational field itself must also participate in
gravitational interaction. These are the two properties that drive us from Newton to Einstein. It is therefore pertinent
to see how are these features actually incorporated in the Einstein's theory of gravitation - general relativity (GR)? \\

The usual derivation of the Schwarzschild solution describing the gravitational field of a mass point in the textbooks does not bring out explicitly these very subtle and important conceptual
aspects. Our main aim in this pedagogical discussion is to demonstrate how these features are so beautifully and elegantly incorporated in GR. In the next
section we shall first discuss a simple derivation of the Schwarzshild solution by demanding that the timelike radial geodesic should include
the Newtonian gravitational law and the velocity of light should remain constant in the empty space surrounding the mass point. It is
interesting that these two simple considerations determine the Schwarzshild solution exactly. Then we solve the vacuum equation and in the
process we expose where and how gravitational interaction of massless particles and self interaction are actually incorporated? We also consider particle orbits again to illuminate some subtle and insightful points. We end up with a discussion. \\

\section{The Schwarszchild field}

For the gravitational field of a mass point which is static and radially symmetry, we begin with the usual spherically symmetric metric,
\begin{equation}
ds^2 = Adt^2 - Bdr^2 - r^2(d\theta^2 + sin^2\theta d\phi^2)
\end{equation}
where $A$ and $B$ are functions of $r$.\\

\subsection{Geodesics}

If GR has to include the Newtonian gravity, the timelike radial geodesic should reduce to $\ddot{r} = -\Phi^{\prime}$ where prime denotes
derivative relative to $r$ . Second if the velocity of light has to remain constant in empty space surrounding the mass point, photon should
experience no acceleration. Since the metric is free of $t$, we immediately write
\begin{equation}
A\dot{t} = E
\end{equation}
where $\dot{r} = dr/ds$ for the timelike particle and $\dot{r} = dr/d\lambda$ for photon. Substituting this in the metric, we get for photon,
\begin{equation}
\ddot{r} = (\frac{E^2}{AB})^{\prime}.
\end{equation}
Now it should experience no acceleration, $\ddot{r} = 0$, means $AB = const.$. Since at infinity, the metric should go over to the flat
Minkowski, hence $AB = 1$. On the other hand for the radially falling timelike particle, we similarly write
\begin{equation}
\dot{r}^2 = \frac{E^2}{AB} - \frac{1}{B} = E^2 - A
\end{equation}
which on differentiation gives
\begin{equation}
\ddot{r} = -\frac{A^{\prime}}{2}.
\end{equation} \\

For GR to include the Newtonian law, $A = 1 + 2\Phi$ with $\Phi$ as the Newtonian gravitational potential. We have thus obtained both the metric
functions, $A = 1/B = 1 + 2\Phi$, which exactly agree with the Schwarzschild solution obtained by solving the non-linear vacuum equation,
$R_{ab} = 0$. This is the simplest derivation of the solution which is purely driven by the physically reasonable considerations of the
inclusion of the Newtonian law and the velocity of light being constant. Note that it is the photon motion which requires space to be curved ($B
\neq 1$) while the Newtonian law would be included for the timelike particle even when space is flat with $B = 1$. It is reflection of the
fact that photon or light can feel gravity only through curvature of space. That is where it freely propagates. It is therefore clear
that Einstein is Newton with space curved. \\

\subsection{Solving the equation}

For the above metric, we have now to solve the vacuum equation,
\begin{equation}
R_{ab} = 0.
\end{equation}
There are three independent components of the Ricci curvature and two of which read as
\begin{equation}
R^t_t = \frac{1}{2AB}[A^{\prime\prime} - \frac{A^{\prime}}{2}(\frac{A^{\prime}}{A} + \frac{B^{\prime}}{B}) + \frac{2A^{\prime}}{r}],
\end{equation}
\begin{equation}
R^r_r = R^t_t + \frac{1}{rB}(\frac{A^{\prime}}{A} + \frac{B^{\prime}}{B}).
\end{equation}
Clearly $R^t_t = R^r_r$ implies $AB = const. = 1$ for asymptotically reducing to the Minkowski flat spacetime \cite{dad1,jacob}. Note that it is
the same condition which followed from photon experiencing no acceleration. This is the condition what is known as the null energy condition
given by $R_{ab}k^ak^b = 0, k_ak^a = 0$. Then writing $A = 1 + 2\Phi$, $R^t_t = 0$ reduces to the familiar Laplace equation \cite{dad2},
\begin{equation}
\nabla^2\Phi = 0
\end{equation}
which integrates to give the familiar solution
\begin{equation}
\Phi = k - M/r.
\end{equation}
Now the remaining equation takes the form
\begin{equation}
R^\theta_\theta = -\frac{2}{r^2}(r\Phi)^{\prime} = 0
\end{equation}
which sets the constant $k=0$. That is, the potential can have zero only at infinity nowhere else. This is in contrast to the Newtonian theory
where the constant $k$ remains free and can be chosen arbitrarily. We have thus obtained the Schwarzshild solution by solving the vacuum
equation $R_{ab} = 0$.

It however raises couple of very interesting questions. First and foremost, where has the self interaction of gravity gone which is the defining
property of the Einsteinian gravity and second, how is it that potential is determined absolutely, vanishing at infinity and nowhere else?
This is what we take up in the next section.\\

\section{Self interaction and zero of the potential}

The new features that Einstein gravity brings in are essentially the two, self interaction and photon feeling gravity. It is therefore
reasonable to expect that the former should facilitate the latter. That is gravitational effect of the self interaction should be such that it
makes photon feel gravity. For photon to feel gravity space has to be curved such that it does not have to change its velocity. This means the
contribution of self interaction is to curve the space without producing any acceleration like $\nabla\Phi$. That is why the Laplace equation
giving the Newton's inverse square law remains intact. This is how the self interaction is incorporated through the curvature of space while the
potential is still given by the good old Laplace equation. \\

The condition for photon to feel no acceleration like ordinary timelike particles is $A^{\prime}/A + B^{\prime}/B = 0$. It is this condition
that reduces $R^t_t = 0$ to the Laplace equation of the Newtonian gravity. If space were flat which means $B = 1$, then it would have taken
the form,
\begin{equation}
\nabla^2\Phi \approx {\Phi^{\prime}}^2
\end{equation}
indicating the self interaction contribution as square of the field, $\Phi^{\prime}$. What really happens is most remarkable that self
interaction contribution goes into curving the space with $B\neq1$ and further the velocity of light should remain constant determines $B =
1/A$. Thus note that the gravitational field energy gravitates in subtler way than matter density which produces $\nabla\Phi$ by curving 
space and not by producing acceleration. This is how it should be because gravitational field energy is not the primary source of gravity
like matter density. It is produced by matter density and has no independent existence of its own. It is a secondary source and hence it should
not do what matter does and sit on the right in the above equation like the matter density. On the other hand for photon not to accelerate but
yet to feel gravity, space must curve and that is precisely what the self interaction does. Thus gravitational field gravitates by curving the
space
without making any contribution to acceleration. This is how self interaction is beautifully incorporated in GR by enlarging the spacetime background and not by modifying the gravitational law \cite{dad5}. \\

The next question is why is the potential determined absolutely, it can vanish only at infinity and nowhere else. In the Newtonian gravity,
potential is determined only up to addition of a constant which can be chosen arbitrarily. In contrast, as we have seen above that the equation
corresponding to $R^\theta_\theta = 0$ determines this constant to be zero leaving no choice for choosing zero of the potential. That is
constant potential attains non-trivial physical meaning here as it produces Ricci curvature $R^\theta_\theta = -2k/r^2$. This is very strange
because in all classical physics constant potential is dynamically trivial and has no physical significance. Let us then ask what is it that is
different for the Einsteinian gravity? It is universal and hence it makes an unusual demand on spacetime that it has to curve to describe its
dynamics. No other force makes such a demand on spacetime. For the rest of the physics, spacetime background is fixed and it is not affected by
the physics happening in it. In contrast, Einsteinian gravitational dynamics can only be described by the spacetime curvature and hence it
cannot remain inert and fixed as for the rest of physical phenomena. Note that in GR, gravitational field is self interactive which means it has
gravitational charge that is spread all over the space up to infinity. So gravitational source is not entirely localized at the location of the
mass point but is distributed all over space. It is a different matter that this distributed source in the form of gravitational field energy
gravitates differently from the mass point but it is nevertheless self interactive source of Einsteinian gravity. Therefore for the Einsteinian
gravitational potential as it occurs in the Schwarzschild solution, space surrounding the mass point is not completely free of "gravitational
source or charge". That is why it cannot vanish in the region which is not completely free of gravitational charge and therefore it can vanish
at infinity and nowhere else. Thus potential in the standard Schwarzschild coordinates gets determined absolutely. \\

\section{Particle orbits and self interaction}

As we saw in Sec.II, space curvature has no effect on radial motion as the equation (5) entirely agrees with the Newtonian law except for derivative here being w.r.t. proper time. It easily integrates to give the finite proper time of fall from radius $r_0$ to $r=0$ as $\sqrt{2r_0^3/9m}$. This is because the inverse square law remains intact and the space curvature does not affect radial motion. It would however make contribution for the non-radial motion. \\

Since the field is radially symmetric, there is no loss of generality in setting $\theta = \pi/2$ and like energy there is also conservation of angular momentum,
\begin{equation}
 r^2 \dot{\phi} = l.
\end{equation}

Substituting the two constants of motion in the metric, we write the standrad expression 
\begin{equation}
 {\dot{r}}^2 = E^2 - (1 - \frac{2m}{r})(\frac{l^2}{r^2} + \mu)
\end{equation}
where $\mu = 1, 0$ refers respectively to timelike and null particle. By differenting the above equation, we write the condition for circular orbit as 
\begin{equation} 
\frac{m}{r^2} + \frac{3ml^2}{r^4} - \frac{l^2}{r^3} = \frac{m}{r^2} - \frac{l^2}{r^3}(1 - \frac{3m}{r}) = 0.  
\end{equation}
Here the first and last terms are the familiar inverse square attarction and the centrifugal repulsion while the middle one is due to space curvature which couples transverse motion with the gravitational potential. By clubbing it with the centrifugal term, it has also been argued \cite{abra} that centrifugal force changes sign at $r=3m$. That is why there cannot exist any circular orbit below this radius. This is also the radius at which occurs the photon circular orbit. Clearly no particle can have circular orbit below the photon orbit radius. Note that the middle term is attractive and is in tune with the first and it is caused by the self interaction. It is gravitational in character and not kinematical and hence it should not be clubbed with the repulsive centrifugal term. Since it produces space curvature which affects transverse motion, that is how it  gets linked to angular momentum. We would rather like to understand the above condition emerging from a potential 
\begin{equation}
 \frac{-2m}{r}(\mu + \frac{l^2}{r^2}) + \frac{l^2}{r^2}
\end{equation}
where the self interaction potential is coupling of gravitational potential with the transverse kinetic energy. Since photon feels no usual $m/r^2$ attraction, it has circular orbit when gradient of the self interaction term balances the centrifugal force. That is why we should not club the self interaction term with the centrifugal force else photon will  have circular orbit with vanishing centrifugal force. Circular orbit is defined by the  balance between attractive and repulsive effects. Effectively space curvature manifests in providing an additional attractive potential for transverse motion. The photon orbit marks the balance between the gradient of this and the centrifugal force. \\

We can in the standard way write the orbit equation for timelike particle, 
\begin{equation}
 u^{\prime\prime} + u = \frac{m}{l^2} + 3mu^2
\end{equation}
which for photon reduces to 
\begin{equation}
u^{\prime\prime} + u = 3mu^2
\end{equation}
where $u=1/r, u^{\prime}=du/d\phi$. Note that it is $3mu^2$ which is the non-Newtonian contribution due to self interaction and it manifests in curving the space. It is clear that photons only feel space curvature. For timelike particles like planets, the orbit would be elliptical in the first approximation because the garvitational attraction law is the same inverse square law. Since the force law is not changed, the nature of orbit has essentially to remain undisturbed. It could then accommodate the effect of space curvature by suffering precession of perihelion. Why perihelion because that is where the force is strongest. Thus self interaction through space curvature make perihelion of the orbit precess. The orbits in the Einstein gravity are therefore precessing ellipses. Further note that gravitational field energy which is negative for positive mass curves space in such a way that it is in consonance with the attraction due to mass. It has been argued elsewhere \cite{dad4} that positive energy condition for gravitational field energy is that it is negative. It defines the norm of positivity for non-localizable energy distribution. For example, the electric field energy of a charged source is positive which is opposite of the norm set by negative gravitational field energy. It is therefore gravitationally 'negative' and that is why it contributes a repulsive effect opposing attraction due to mass for the field of a charged black hole. Let us consider potential at some $r$ due to a charged particle of mass M and charge Q which would be given by 
\begin{equation}
\Phi = - \frac{M - Q^2/2r}{r}. 
\end{equation}
This is because the electric field energy, $Q^2/2r$, lying outside the radius $r$ does not contribute and hence has to be subtracted out. It would give rise to the acceleration $-M/r^2 + Q^2/r^3$ which shows the repulsive effect of the electric field energy \cite{dad7}. Since electric field energy is positive, it is therefore ``gravitationally negative'' and hence repulsive. \\  

\section{Discussion}

The main aim of this note is to bring out transparently inclusion of gravitational self interaction and its role in particle orbits, and why potential in the Schwarzschild
solution cannot vanish anywhere but at infinity. This is very interesting and insightful for appreciating the remarkable features of the
Einsteinian gravity over the Newtonian gravity. Note that ultimately the equation we need to solve is the first order linear differential
equation which is the first integral of the Laplace equation. It is this that determines the potential absolutely. The Newtonian $\nabla\Phi$ comes
from $A = 1 + 2\Phi$ which can be squared out by redefining $t$ when $\Phi = k$. It is $A$ that gives the Newtonian acceleration, $\nabla\Phi$
and hence $A=const$ has no physical significance. However $\Phi = k$ in $B$ has non-trivial effect because it refers to curvature of space which
is sourced by self interaction and it does not vanish when $B = const.$. It may be noted that the constant potential generates the following
stresses,
\begin{equation}
T^t_t = T^r_r = \frac{k}{r^2}, \, \, \, \, T^\theta_\theta = T^\phi_\phi = 0
\end{equation}
and they asymptotically agree with that of a global monopole \cite{bv,dad3}. It is remarkable that constant potential dynamically therefore
describes a global monopole.

The Einsteinian gravity is essentially driven from the Newtonian gravity by the two new properties of self interaction and photon feeling
gravity without experiencing acceleration. The former contributes through curving space which also facilitates photon's interaction with
gravity. It is remarkable that the two new properties are intimately related to each-other leaving essentially the Newtonian gravity intact. The
standard derivation and discussion of the Schwarzschild solution do not expose these interesting new aspects of the Einsteinian gravity in such
a transparent and explicit manner. That is precisely what we had set out to do. \\

{\it Acknowledgment:} It is a pleasure to thank the Al Faraby Kazakh National University, Almaty for the kind invitation to lecture in the
School on Theoretical Physics organized by the Department of Theoretical and Nuclear Physics. These subtle aspects, which are otherwise not so
explicitly and transparently discussed, were brought out succinctly in the author's lectures. I warmly thank Professor Medeu Abishev for the
wonderful hospitality.

%======================================%
%<<<<<<<<<<<<< REFERENCES >>>>>>>>>>>>>%
%======================================%


\begin{thebibliography}{9}
\bibitem{dad5}
The vacuum energy arising from the quantum fluctuations is on the same footing as the gravitational field energy and has no independent existence of its own. It should also therefore gravitate not like other matter through a stress tensor but rather subtly in the same manner as the gravitational field energy \cite{dad6}.
\end{thebibliography}
\end{document}